\input  phyzzx
\input epsf
\overfullrule=0pt
\hsize=6.5truein
\vsize=9.0truein
\voffset=-0.1truein
\hoffset=-0.1truein

%
%
\def\half{{1\over 2}}
\def\IC{{\ \hbox{{\rm I}\kern-.6em\hbox{\bf C}}}}
\def\IR{{\hbox{{\rm I}\kern-.2em\hbox{\rm R}}}}
\def\IZ{{\hbox{{\rm Z}\kern-.4em\hbox{\rm Z}}}}

\def\sIR{{\hbox{{\sevenrm I}\kern-.2em\hbox{\sevenrm R}}}}

\def\sym{super Yang Mills theory}

\def\mt{Matrix theory}

\def\n.{{N \over 2}}
%
%
\hyphenation{Min-kow-ski}
\rightline{UTTG-18-97}
\rightline{RU-39}
\rightline{SU-ITP-97-28}
\rightline{hep-th/9705190}
\rightline{May 1997}

\vfill

%
%
\title{Instantons, Scale Invariance and Lorentz Invariance in  
Matrix Theory}

\vfill

%
%
\author{Tom Banks$^1$,
W.Fischler$^2$,
Nathan Seiberg$^1$
 and
L. Susskind$^3$}

\vfill

\address{$^1$Department of Physics and Astronomy,\break Rutgers  
University,
Piscataway, NJ 08855-0849}
\address{$^2$Theory Group,Department of Physics,\break University  
of Texas,
Austin,TX 78712}
\address{$^3$Department of Physics,  Stanford University\break  
Stanford, CA
94305-4060}
\vfill


\vfill

\vfill

%
%
In this paper we consider features of graviton scattering in Matrix
theory compactified on a 2-torus. The features which interest us
 can only be determined by nonperturbative effects in the
corresponding 2+1 dimensional super Yang Mills theory. We show that
the superconformal symmetry of strongly coupled Super Yang Mills
Theory in $2+1$ dimensions almost determines low energy, large impact
parameter ten dimensional graviton scattering at zero longitudinal
momentum in the Matrix model of $IIB$ string theory.  We then show
that amplitudes involving arbitrary transverse momentum transfer are
governed by instanton processes similar to the Polchinski Pouliot
process.  Finally we consider the influence of instantons on a
conjectured nonrenormalization theorem.  This theorem is violated by
instanton processes. Far from being a problem, this fact is seen to be
crucial to the consistency of the $IIB$ interpretation.
We suggest that the $SO(8)$ invariance of strongly coupled SYM theory
may lead to a proof of eleven dimensional Lorentz invariance.
\vfill\endpage

%
%
\REF\aspsch{P. Aspinwall, Talk Presented at ICTP Trieste Conf. on
Physical and Mathematical Implications of Mirror Symmetry in String
Theory, June 1995, hep-th/9508154; J. Schwarz, 
Phys. Lett.{\bf B367}(1996)97, hep-th/9510086.} 
\REF\bfss{T. Banks,W. Fischler, S. Shenker and L.  
Susskind, hep-th/9610043.}
\REF\wati{W.Taylor, hep-th/9611042.}
\REF\grt{O.J. Ganor, S. Ramgoolam and W.Taylor, hep-th/9611202.}
\REF\bss{T. Banks, N. Seiberg and S. Shenker, Nucl.Phys. {\bf B490}
(1997) 91, hep-th/9612157.}
\REF\ss{S. Sethi and L. Susskind, hep-th/9702101.}
\REF\bs{T. Banks and N. Seiberg, hep-th/9702187.}
\REF\pp{J. Polchinski and P. Pouliot, hep-th/9704029}
\REF\dkm{N. Dorey, V.V. Khoze and M.P. Mattis, hep-th/9704197}
\REF\seib{N. Seiberg, hep-th/9705117.}
\REF\tb{T. Banks, Trieste Lectures 1997 to appear}
\REF\fhrs{W. Fischler, E. Halyo, A. Rajaraman and L. Susskind,  
hep-th/9703102.}
\REF\ds{M. Dine and N. Seiberg, hep-th/9705057}
\REF\bb{K. Becker and M. Becker, hep-th/9705091}

%
%

%
%
\chapter{Introduction}

The emergence of spatial dimensions in string theory often occurs
through mechanisms which seem extremely bizarre from any conventional
viewpoint.  Two notable examples are the emergence of the 11th
direction of M-theory in strongly coupled IIA string theory and the
10th direction of type $IIB$ theory which materializes when M-theory
is compactified on a 2-torus of vanishing area [\aspsch].  In both
cases a discrete quantum number is identified which then plays the
role of momentum for the new direction. In the first case the discrete
quantum number is the number of D0-branes. In Matrix theory this
becomes $N$, the rank of the $U(N)$ gauge group. As discussed in
[\bfss], $N$ is identified with the longitudinal momentum of the light
cone description.  In the small 2-torus case the 2-brane wrapping
number is the conjugate momentum to the new direction which we call
$Y$.  This quantum number also has significance in Matrix theory.
Matrix theory on a 2-torus is described by 2+1 dimensional $U(N)$
\sym\ with 16 supersymmetries [\bfss, \wati]. The 2-brane wrapping
number becomes the magnetic flux through the torus [\grt\ -- \bs]. 

In both of the above cases the symmetry which rotates these new
directions into the other directions is highly non-manifest. In the
first case the symmetry is the difficult ``angular conditions'' which
rotate longitudinal into transverse directions. In the second case
they are the transverse rotations which rotate the new Aspinwall-Schwarz
direction into
the other $7$ transverse directions. In each case there is evidence
for the exactness of these symmetries in appropriate limits; the large
$N$ limit in one case and the large membrane wrapping number in
the other. These limits generally involve letting the discrete quantum
number tend to infinity while the corresponding quantum of energy
tends to zero. In the second case we allow the torus to shrink to zero
area while increasing the integer valued magnetic flux.

Evidence for the restoration of Lorentz invariance in the large $N$
limit of Matrix theory was given in [\bfss] where it was shown that
low energy graviton-graviton scattering with zero longitudinal
momentum transfer has exactly the form of single graviton
exchange. However, to establish the full invariance, it is necessary
to gain control over processes in which longitudinal momentum is
exchanged. Progress has been made on this problem by Polchinski and
Pouliot [\pp], who studied processes in which longitudinal momentum is
transferred between infinite membrane configurations of Matrix
theory. Here the basic process of unit momentum transfer was found to
correspond to an instanton in a 2+1 dimensional gauge theory
describing the membranes. The instanton amplitude exactly agrees with
the expected behavior from single graviton exchange. Further progress
was made by Dorey, Khoze and Mattis [\dkm], who were able to study the
sum over instantons, allowing any number of momentum units to be
exchanged. At the moment it has not yet been possible to study
longitudinal momentum exchange between types of configurations other
than these extended membranes. For example one would also like to
exchange longitudinal momentum between gravitons.

In the case of the vanishing torus and type $IIB$ theory two
arguments have been given for rotational invariance [\ss, \bs,
\seib].  Electric-magnetic duality of 3+1 dimensional \sym \ was used
to indirectly prove that the manifest $SO(7)$ invariance of the \sym \
is enhanced to $SO(8)$ in the limit of vanishing torus area. An
alternate argument is based on the fact that the vanishing torus limit
should be described by a 2+1 dimensional fixed point theory.  Then the
superconformal invariance requires the $SO(8)$ symmetry. We will see
in the next section that the superconformal invariance also determines
a great deal about how graviton scattering amplitudes depends on
impact parameter.

In this paper we will be interested in scattering amplitudes of
gravitons in which $Y$ momentum is exchanged. For graviton-graviton
scattering amplitudes involving no exchange of either longitudinal
momentum or momentum in the $Y$ direction the method of [\bfss] shows
that the scattering is described by single graviton exchange in Matrix
theory. However, to demonstrate the rotational invariance which
rotates $Y$ into the other noncompact directions it is necessary to
exchange $P_Y$. We shall see in what follows that this process is
again governed by the same 2+1 dimensional instanton
amplitudes as in [\pp].

Let us define some notations.  The size of the two cycles of the torus
are $L_1,L_2$.  The longitudinal direction of the infinite momentum
description of M-theory will be denoted by $x^{11}$. Following [\bfss]
we take $x^{11}$ to be compact with radius $R$. The large $N$ limit
effectively decompactifies $x^{11}$. The 9 transverse directions are
$X^i$ with $i=1,...,9$. The two compact directions of the 2-torus are
$X^1,X^2$. This leaves 7 manifest transverse spatial
directions. Finally the newly emergent transverse direction is called
$Y$. We can also denote this direction by $X^{10}$ but we prefer a
notation which emphasizes the asymmetry of the M-theoretic origin of
$Y$. Finally the term ``longitudinal'' momentum always refers to the
component of momentum along the longitudinal axis of the light cone
frame.

To conclude this introduction we will make some general remarks about
scattering in Matrix theory. The definition of scattering amplitudes
in Matrix theory involves path integrals with fixed boundary
conditions on the moduli space in the asymptotic past and
future\foot{See [\tb] for a more detailed discussion.}.  As in the
LSZ formalism of field theory, it is only necessary to choose boundary
conditions which have a finite overlap with some stable state of the
model.  Scattering amplitudes are extracted from these path integrals
by dividing by a wave function renormalization factor which extracts
the overlap between the state defined by the asymptotic boundary
conditions, and the true scattering states of the system.

Now consider the scattering of two graviton states of the toroidally
compactified matrix model, with zero longitudinal momentum exchange
and very large transverse separation.  Even though we do not know the
wave function of these states for generic values of the $IIB$ string
coupling it is extremely plausible that this can be encoded in an
effective Lagrangian for the coordinate describing the separation
between their centers of mass \foot{In local field theory this is a
theorem and can be proven by dispersion relations.  In the present
context one may be suspicious of it because of the growth of wave
functions with $N$ described in [\bfss].  In particular, Steve
Shenker has reminded us that in the stringy regime, $N$ (which
controls the world sheet cutoff) must always be taken large enough for
the perturbative string wave functions to overlap, in order to
reproduce long range gravitational interactions. However, in the
presence of maximal SUSY, it appears that one can obtain correct
results without taking the large $N$ limit.}.  On the moduli space,
this coordinate is the zero mode of a field $\Delta^a , \quad a = 1
\ldots 8$, which is the difference between the coefficients of the
unit matrices in the two blocks representing the particles.

Supersymmetry guarantees that any correction to the free motion on the
moduli space must contain at least four derivatives (or various powers
of fermions).  Thus, we may expect a Lagrangian of the schematic form
$${\delta {\cal L} = F_{abcd} (\Delta^a ) \partial \Delta^a
\partial \Delta ^b
\partial \Delta^c \partial \Delta^d}
\eqn\oneone
$$
Calculating low energy scattering amplitudes between gravitons requires
computing the functions $ F_{abcd}$.

%
%
\chapter{Instantons and Momentum Transfer in $IIB$ Theory}

The  2+1 dimensional \sym \ describing
$IIB$ theory lives on the  torus dual to $L_1,L_2$ with sides
$\Sigma_1,\Sigma_2$ [\fhrs].
$$
\Sigma_i=(2\pi)^2{l_{11}^3 \over L_iR}
\eqn\twoone
$$
where $l_{11}$ is the 11 dimensional Planck length and $R$ is the
compactification radius of the longitudinal direction [\bfss].

 The field content is seven scalars, $\phi$, a vector $A$ and fermions
$\psi$. The  Yang Mills coupling constant $g$ is given by [\fhrs]
$$
g^2=(2\pi)^2{R \over L_1L_2}.
\eqn\twotwo
$$
The fields $\phi$ are proportional to the 7 uncompactified transverse
coordinates
$X^3, X^4,....X^9$. The precise connection is
$$
\phi={gX \over \sqrt{R \Sigma_1 \Sigma_2}}.
\eqn\twothree
$$
The $U(1)$ magnetic flux through the torus $\Sigma_{1,2}$ is quantized
in integer units.
The energy of a single unit of flux is given by
$$
E_M ={(2\pi)^2 \over 2Ng^2\Sigma_1 \Sigma_2} .
\eqn\twofour
$$
It can be reexpressed in terms of the $L_i$, $R$ and $l_{11}$
$$
E_M = {R\over 2N}{L_1^2 L_2^2 \over (2\pi)^4l_{11}^6}={1\over  
2p_{11}}{L_1^2 L_2^2 \over (2\pi)^4l_{11}^6}.
\eqn\twofive
$$
{}From \twofive\ it is seen that the energy gap tends to zero as the
torus $L_{1,2}$ shrinks. This is properly interpreted as the  
decompactification of
$Y$. By matching energy scales it was shown in [\fhrs] that the size of
the
$Y$ circle  $L_Y$, satisfies
$$
L_YL_1L_2=(2\pi)^3l_{11}^3 .
\eqn\twosix
$$

The $Y$ component of momentum $P_Y$ is related to the integer valued
magnetic flux quantum number $n$ by
$$
P_Y = 2 \pi {n \over L_Y}
\eqn\twoseven
$$

We will consider the scattering of a pair of gravitons in the $IIB$ theory.
For simplicity
we take the longitudinal momenta of the gravitons to be equal. This
condition may
be relaxed without introducing any essential complication. The  
configuration is
described in \mt \ by considering block diagonal matrices composed  
of equal
size
blocks $\n.
\times \n.$
blocks. The center of masses of the two blocks are well separated in the
transverse 7
dimensional space. The separation 7-vector has length $\rho$ which
corresponds to a
symmetry breaking field
$$
|\phi| = \rho {g \over \sqrt {R \Sigma_1 \Sigma_2}}
\eqn\twoeight
$$
 This corresponds to spontaneously breaking the
$U(N)$ gauge group to
$U(\n.)
\times U(\n.)$. We will assume that all momentum invariants are small and
that the impact parameter is large.

The spatial momentum transfer in the scattering process is described
by a nine-vector $Q$ with components in the seven directions $X$, the
longitudinal direction $x^{11}$ and the $Y$ direction. For processes
with $Q _{11} = Q _Y =0$, the amplitude is given by the same methods
as used in [\bfss]. Recall that when the $U(N)$ symmetry is broken,
the strings connecting the two blocks become massive. Integrating out
these degrees of freedom in the one loop approximation gives a
graviton-graviton scattering amplitude which agrees with single
graviton exchange. The thing we wish to emphasize is that the
amplitude in this case is computed by a purely perturbative method in
the \sym.

We will continue to consider vanishing $Q_{11}$ but relax the
condition $Q_Y =0$. As we shall see such processes are nonperturbative
in the Yang Mills coupling $g$. With no loss of generality we may
suppose that the two initial gravitons have equal and opposite
transverse momentum including the component $P_Y$. The momenta $P_Y$
of the two gravitons are described as follows. When the gauge group is
broken to $U(\n.) \times U(\n.)$.  There are two Abelian $U(1)$
magnetic fluxes $n_1,n_2$ associated with the two blocks. Choosing
$n_1 = -n_2 =n$, the initial $Y$ momenta are $ \pm {n \over L_Y}$. Now
consider a process in which $n_1 \to n_1 +1$ and $n_2 \to n_2 -1$ or
equivalently $n \to n+1$. This corresponds to a momentum transfer with
$Q_Y = {2 \pi \over L_Y}$.  We may also think of it in gauge theory
language. Consider an $SU(2)$ subgroup of $U(N)$ between a zero brane
in each group. For example the $SU(2)$ which mixes the entries labeled
$1$ and ${N\over2}+1$.  Its generators are Pauli matrices $\tau$
acting in the $2 \times 2$ subspace $(1,{N\over2}+1)$. This group is
broken to the $U(1)$ generated by $\tau_3 $. The transition which
exchanges $Q_Y={2 \pi\over L_Y}$ may be thought of as changing the
magnetic flux of this $U(1)$ subgroup by one quantum.

A process in which $SU(2)$ magnetic flux changes by an integer flux
quantum is topologically nontrivial and requires a nonvanishing
instanton number. The process is almost mathematically identical to
the process studied in [\pp]. There are however some differences:

\item {1.} Polchinski and
Pouliot were considering the scattering of infinite 2-branes which
exchange longitudinal momentum. The 2+1 dimensional field theory
describing the branes lives on an infinite plane. In the present case
the 2+1 dimensional field theory lives on the torus dual to the
space-time torus.

\item {2.} In [\pp] the gauge group corresponding to the two 2-branes was
$U(2)$. In our case the system carries $N$ units of longitudinal
momentum (not to be confused with $P_Y$) and is described by the gauge
group $U(N)$.

However the difference between the two cases does not lead to
substantial differences in the calculation. First, in the $IIB$ limit
in which $L_Y \to \infty$ the field theory torus tends to infinite
size while the instantons of interest remain small.  Second, the
instantons live in $SU(2)$ subspaces. Integrating over the orientation
of the instantons only affects the results by multiplying the $n$
instanton result by an $N$ dependent factor $C_N$

Let us first estimate the instanton amplitude for large values of the
separation of the two gravitons (large Higgs field). The action of an
$SU(2)$ instanton is given by
$$
S={2 \pi \over g^2}f
\eqn\twonine
$$
where $f$ is the expectation value of $\phi$ in the broken symmetry
state. Using \twoone, \twotwo \ and \twothree \ we find
$$
S={L_1L_2 \rho \over 2 \pi^2 l_{11}^3}
\eqn\twoten
$$
where  $\rho$ represents the 7 dimensional distance between the
gravitons.  Finally, \twosix \ gives
$$
S={2\pi \over L_Y} \rho
\eqn\twoeleven
$$
This means that the amplitude for the exchange of a single quantized unit
of $P_Y$
is
$$
A=\exp \left[ -{2\pi \over L_Y} \rho \right ].
$$

It would be interesting to complete the instanton computation for gauge
group $U(N)$ for any $N$ and finite volume and to perform the sum over
instanton numbers along the lines of [\dkm].  This computation is not
easy. However, as $L_Y \to
\infty$ the field theory torus tends to infinite size and we may carry
out our calculations in infinite space. In this limit the effective
Lagrangian \oneone \ (with zero modes replaced by local fields)
describes a superconformal fixed point theory with scale invariance
and $SO(8)$ symmetry [\seib]. We may use the scale invariance to
constrain the functional form of the $F_{abcd}(r)$.

To do so we note that out along the flat directions the fields $X$
have scaling dimension $\half$. This is determined by the form of the
quadratic term in ${\cal L}$. The translation invariance of the system
requires this term to be the usual canonical free field Lagrangian
{}from which the dimensionality of $X$ follows.  We then determine the
dimensionality of $F_{abcd}(r)$ to be 3. This means that $F$ must be
proportional to $ r^{-6}$.

In fact, this scale invariance argument suggests that the full answer
is not just proportional to $v^4 \over r^6$ but there can be three
terms 
$$
A{(v \cdot v)^2 \over r^6} + B {(v \cdot v) (r \cdot v)^2 \over r^8} +
C {(r \cdot v)^4 \over r^{10} }
\eqn\newequ
$$
where $A,B,C$ are coefficients which cannot be determined by using the
$SO(8)$ and scale symmetries alone.  It is likely that the ratios
between them can be determined using supersymmetry.

%
\chapter{Conclusion}

When Matrix theory is compactified on a 2-torus a new direction of
space which we called $Y$ emerges as the torus shrinks to zero
size. We have seen that graviton scattering processes involving the
exchange of the $Y$ component of momentum are nonperturbative
instanton processes in the 2+1 dimensional \sym \ describing the
compactified \mt .  Using the calculations of [\pp] and [\dkm] we saw
that the scattering amplitude for such processes agrees with the
results of supergravity perturbation theory when the impact parameter
is bigger than $L_Y$, the compactification scale of $Y$. However if
one wants to study the theory for fixed impact parameter as $L_Y \to
\infty$ one has to go beyond the leading order semiclassical instanton
approximation. This may be done by appealing to the superconformal
invariance of the fixed point theory describing the strongly coupled
theory. The result derived in section 2 agrees with supergravity
calculations at all length scales between $L_Y$ and the 10 dimensional
Planck scale. In particular, the amplitude is of the form ${{v}^4 /
r^6}$ and is invariant under the $SO(8)$ rotation group.

According to Polchinski and Pouliot [\pp] the same \sym \ which
describes the toroidal compactification also describes the theory of
2-branes in the uncompactified theory. This theory may be used to
describe longitudinal momentum transfer in the scattering of infinite
2-branes. One may wonder what the implications of the superconformal
fixed point theory are for the membrane amplitudes. Let us suppose
that we have a collection of any number $k$ of parallel membranes
oriented in the $X^1,X^2$ plane.  In the Polchinski, Pouliot
description of Matrix theory this is described by $U(k)$ \sym.  The
membranes may not all be at rest in the same frame. In other words
they may be relatively boosted along $x^{11}$ with respect to one
another. This is described by turning on various magnetic fluxes
associated with the unbroken $U(1)$ subgroups which survive when the
branes are separated.

Polchinski and Pouliot find that the \sym \ becomes strongly coupled
in the limit $R \to \infty$. Therefore when $x^{11}$ decompactifies
the membrane interactions and scattering are described by the strongly
coupled fixed point theory which is $SO(8)$ invariant. The $SO(8)$
group in this case is just the rotation group acting on the 8 spatial
dimensions of the 11 dimensional theory which are transverse to the
branes $including $ $x^{11}$.

What does this say about Lorentz invariance of supergraviton
scattering in 11 dimensions? To answer this we may use the fact that
the poles in scattering amplitudes factorize.  If for each external
supergraviton in a process, we introduce a 2-brane to act as a source,
the $SO(8)$ invariance of the membrane amplitude guarantees the
corresponding invariance of the supergraviton scattering. This
together with the manifest $SO(9)$ invariance of the Matrix theory of
supergravitons should provide a basis for a proof of 11 dimensional
Lorentz invariance.

We will conclude with some remarks about the nonperturbative breakdown
of the nonrenormalization theorem reported in [\ds]. Let us recall the
argument in [\bfss] for the necessity of such a theorem. In that
reference a one loop matrix quantum mechanics calculation of the force
between two gravitons in 11 dimensions was shown to exactly agree with
supergravity at large distances and small transverse momentum. The
amplitude had the form
$$
A \sim N^2{v^4\over \rho^7}
\eqn\fourone
$$
where the factor $v^4$ is schematic for a quartic expression in the
transverse velocities.  As explained in [\bfss], higher loop
corrections, if they exist, will correct this by a factor of the form
$$
1+c_2  {N \over \rho^3} + c_3\left[{N \over \rho^3}\right]^2  
+c_4\left[{N \over
\rho^3}\right]^3 +... = F\left({N \over \rho^3}\right)
\eqn\fourtwo
$$
Eq.\fourtwo \ represents the leading large $N$ behavior of the loop
diagrams assuming no cancellation takes place. The amplitude will only
agree with graviton exchange if $ F\left({N \over
\rho^3}\right) \to 1$ for $\rho>>l_{11}$. For fixed $N$ \fourtwo \  
shows that this is so. However the correct limit is to fix $\rho$ and
let $N \to \infty$.  Evidently any nontrivial dependence of $F$ is
dangerous in the large $N$ limit.  This circumstance led to the
conjecture that $F$ is not corrected beyond one loop.  The conjecture
has been confirmed at the two loop level [\bb].  Incidentally, it is
obvious from what has been said that the nonrenormalization theorem is
only really required for the leading large $N$ behavior, in other
words for planar diagrams. The calculation in [\bb] does not test this
issue because at the level of two loops the only graphs which
contribute are planar.

Recently, a nonperturbative nonrenormalization theorem of this sort
for 3+1 dimensional \sym \ was proven [\ds].  However, it was also
shown that in 2+1 dimensions instantons violate any such theorem.  In
fact the instanton effects are exactly the ones we have been
discussing in the previous section. The question is whether these
effects are dangerous from the point of view of [\bfss]. To answer
this we note two things. First of all, for finite tori, the instanton
effects are exponentially
suppressed at large distance.  Furthermore, unlike the perturbative
corrections they do not depend on the ratio $N/\rho^3$. They are
perfectly harmless for large $\rho$ as $N$ becomes large.

But as we have seen, the range of these effects grows as $L_Y \to
\infty$. In this limit the long range behavior of the amplitude is
indeed modified by the instanton effects. Far from being a problem,
these effects are essential to describe the emergence of the new
non-manifest $Y$ direction in $IIB$ theory.

\bigskip

\centerline{\bf Acknowledgments}

We would like to thank Joe Polchinski and Steve Shenker for valuable
conversations.  W.F. would like to thank The Stanford Institute for
Theoretical Physics for hospitality while this work was being done.
The work of T.B. and N.S. was supported in part by the Department of
Energy under Grant No. DE - FG02 - 96ER40959 and that of W.F. was
supported in part by the Robert A. Welch Foundation and by NSF Grant
PHY-9511632.  L.S. acknowledges the support of the NSF under Grant
No. PHY - 9219345.

\refout
\end